\documentclass[12pt]{article}
\usepackage{epsf}
\usepackage{epsfig}
\usepackage{graphicx}
\usepackage{amssymb}
\usepackage{bbm}

\def\d{{\mathrm{d}}}

\def\tautheta{{\xi}}

\def\zero{{{\tilde \varphi_0}}}
\def\sign{{{\mathrm{sign}}}}
\def\ff{{h}}
\def\tds{{\xi}}

\date{\today}

\newcommand{\mysection}[1]{\section{#1}
                    \setcounter{table}{0}\setcounter{equation}{0}}
\newcommand{\DD}[1]{{{{{\mathrm{D}\hspace{-0.7em}/\hspace{0.2em}}_{(#1)}}}}}

\def\C{{\mathbb{C}}}
\def\T{{\mathbb{T}}}
\def\R{{\mathbb{R}}}
\def\Z{{\mathbb{Z}}}

\def\L{{\mathbb{L}}}

\def\Id{{\mathbbm{1}}}
\def\tr{{\mathrm{tr}}}
\def\D{{\mathrm{D} }}

\def\ha{\frac{1}{2}}
\def\iha{\frac{i}{2}}

\begin{document}

\begin{titlepage}

\title{Spectral flow of the Dirac spectrum in intersecting 
vortices\footnote{supported by DFG under grant-No. DFG-Re
856/5-1}}

\author{H.~Reinhardt and T.~Tok\\
\vspace{0.5cm}
University of T\"ubingen}

\maketitle

\vspace{2cm}

\normalsize

\begin{abstract} 
The spectrum of the Dirac Hamiltonian in the background of
crossing vortices is studied. To exploit the index theorem, and in
analogy to the lattice the space-time manifold is chosen to be
the four-torus $\T^4$. For sake of simplicity we consider two idealized 
cases: infinitely fat and thin transversally intersecting
vortices. The time-dependent spectrum
of the Dirac Hamiltonian is calculated and in particular the 
influence of the vortex crossing on the quark spectrum is investigated.
For the infinitely fat intersecting vortices it is found that zero modes 
of the four-dimensional Dirac operator can be expressed in terms of 
those eigenspinors of the Euclidean time-dependent Dirac Hamiltonian, which 
cross zero energy. For 
thin intersecting vortices the time 
gradient of the spectral flow of the Dirac Hamiltonian 
is steepest at the time at which the vortices cross each other.
\end{abstract}


\end{titlepage}

\mysection{Introduction}

Center vortices offer an appealing picture of confinement 
\cite{'tHooft:1978hy,Mack:1979rq}, 
for a recent review see  \cite{Greensite:2003bk}. 
Furthermore, the deconfinement phase
transition can be understood as a depercolation transition 
\cite{Engelhardt:1999fd} and the free energy of a center vortex vanishes
in the confinement regime \cite{Kovacs:2000sy}.
In fact there is mounting evidence that center vortices play an important
r\^ole in the QCD vacuum. Lattice calculations show that 
when center vortices are removed from the Yang-Mills ensemble 
the string tension \cite{DelDebbio:1997mh} as well as the quark 
condensate and topological charge \cite{deForcrand:1999ms} are 
lost\footnote{In fact the finding of 
the lattice calculation concerning string tension and deconfinement 
phase transition can be reproduced in a center vortex model 
\cite{Engelhardt:1999wr,Engelhardt:2001ze}.}. 
This indicates that confinement, spontaneous
breaking of chiral symmetry and the chiral anomaly are all triggered by center
vortex configurations in the Yang-Mills vacuum. 
Furthermore, lattice calculations \cite{Karsch:1998ua}
show that restoration of chiral symmetry occurs at finite temperature
exactly at the deconfinement phase transition. We can therefore 
expect that confinement 
and spontaneous breaking of chiral symmetry are related phenomena. While the
emergence of the confining potential, i.e.~the string tension, as well as the
appearance of the deconfinement phase transition can be rather easily 
understood in a random center vortex model in an analytic fashion, the 
mechanism of spontaneous breaking of chiral symmetry has not been explained 
in the vortex picture of the QCD vacuum, yet. The general folklore is that 
spontaneous breaking of chiral symmetry is generated by the quasi-zero 
modes in topologically non-trivial field configurations, which give rise 
to a quasi continuous Dirac spectrum with non-zero level density 
$\rho (0)$ at zero virtuality and, by the Banks Casher relation, to a 
non-zero quark condensate $<q q> \sim \rho (0)$. In fact these calculations 
\cite{Gattringer:2002dv} show, that a non-zero level density at zero virtuality 
occurs in the confinement phase while a finite gap in the Dirac spectrum 
is found above the critical temperature.

It is therefore clear that the topological properties of center vortices and
their manifestation in the quark spectrum will play the key r\^ole in the
understanding of spontaneous breaking of chiral symmetry in the center vortex
picture of the QCD vacuum. In $D=4$ space-time dimensions 
center vortices represent closed surfaces of quantized electromagnetic 
flux which produce a non-trivial center element for a Wilson
loop when they are topologically non-trivially linked to the latter
\cite{Reinhardt:2001kf}. The topological charge of center vortices 
is given by their (oriented) intersection number 
\cite{Reinhardt:2001kf,Engelhardt:1999xw,Cornwall:1999xw}. 
According to the index
theorem a non-zero topological charge implies zero modes of the Dirac
operator \cite{Atiyah:1980jh}. In \cite{Reinhardt:2002cm} the quark 
zero modes in transversally intersecting
vortex surfaces have been studied. It was found that these modes are
localized at the vortex sheets, in particular at their intersection
points. In $D=3$ dimensions a center vortex represents generically a 
time-dependent closed flux loop. In \cite{Reinhardt:2001kf} it was shown 
that the topological charge of generic center vortices can be expressed 
by the temporal change of their writhing number, and furthermore 
transversal intersection points in $D=4$ correspond to crossing of 
vortex lines in $D=3$. Thus vortex crossings carry topological charge
and by the index theorem should consequently manifest themselves in the
quark spectrum.

In this paper we study the quark spectrum in the background of 
time-dependent crossing 
center vortices. In particular we investigate how the crossing 
of vortices manifests itself in the quark spectrum. This might bring 
some light on the mechanism of spontaneous breaking of chiral symmetry 
in the vortex picture, since it is believed that chiral symmetry breaking
is tightly related to the topological properties of gauge fields. 

Throughout this paper the space-time-manifold is chosen to be the 
four-torus $\T^4$. There is a variety of reasons for studying $\T^4$: 
first, $\T^4$ allows to use the Atiyah-Singer index theorem in
a stringent fashion and allows for normalizable spinors in the
background of integer flux. Second, the torus simulates a periodic 
arrangement of vortices. This is much closer to a percolated vortex 
cluster than two isolated vortices in $\R^4$. Finally, the torus 
is the space-time manifold that is also used in lattice calculations.

The quark spectrum and its spectral flow in
time-dependent background gauge fields have been addressed in several
publications before, 
e.g.~\cite{Christ:1980zm,Khoze:1995yb,Klinkhamer:2001cp}. 
In these papers strict relations between the 
spectral flow of the Dirac Hamiltonian and topological properties 
of the corresponding background gauge field 
have been derived in Minkowski space-time.
In \cite{Christ:1980zm,Khoze:1995yb} the spectral flow was considered on
an infinite time interval, where the background gauge field 
has to become a pure gauge at $t = -\infty$ and $t=+\infty$. In 
\cite{Klinkhamer:2001cp} only spherically symmetric gauge fields have
been examined, but the derived results are also valid on a finite time 
interval. In this paper we analyze the spectral flow in the presence of
intersecting center vortices on a compact Euclidean space-time manifold.
The present paper is a continuation of ref.~\cite{Reinhardt:2002cm}
where the quark zero modes in intersecting center vortex sheets were
considered. In the present paper we will not confine ourselves to the
quark zero modes but study the spectral flow as function of the Euclidean
time for the time-dependent vortex crossings. From studying the spectral
flow we hope to get some insight into the mechanism which produces the
quasi-continuous spectrum of Dirac states near zero virtuality. As a
byproduct we will also establish a strict relation between the quark zero
modes and the (time-dependent) eigenfunctions of the Dirac Hamiltonian
in the case where the background fields consist of ``vortices'' with 
homogeneously spread out magnetic flux.
 
The organization of the paper is as follows:
In the next section we consider the mathematical idealization of infinitely
thick transversally intersecting vortices, i.e.~we consider ``vortices''
which are homogeneously distributed over the transversal dimensions, so that
they represent homogeneous field configurations of constant electromagnetic
flux. We consider two orthogonally homogeneously distributed fluxes, which 
can be considered as two infinitely thick intersecting vortices, where the 
intersection ``point'' is smeared out over the whole universe. Accordingly 
the topological charge is homogeneously distributed over space-time. We 
will study then the eigenmodes of the Dirac Hamiltonian as function of 
the adiabatic time-evolution of the vortex flux assuming that the 
three-space is given by a three-torus. In section \ref{orthogonal-thin} 
we study the opposite case of thin transversally 
intersecting vortices assuming that the two intersecting 
vortices are orthogonal to each other. We study the spectral flow of 
the quarks in this vortex background. Finally in section
\ref{non-orthogonal-thin} we allow the intersecting vortex sheets 
to be non-orthogonal and calculate again the spectral flow of the quarks. 
Some concluding remarks are given in section \ref{conclusion}.

\mysection{Fermions in homogeneously intersecting ``vortex'' fields}
\label{orthogonal-fat}

The topological charge of a gauge field configuration 
\begin{eqnarray}
\label{top-charge}
\nu &=& \frac{1}{32 \pi^2} \int \tr F \tilde F
\end{eqnarray}
consisting of center
vortices is located at the intersection points of the vortices
\cite{Engelhardt:1999xw,Reinhardt:2001kf}. In \cite{Reinhardt:2001kf} 
the evolution in time of the topological charge of crossing center 
vortex configuration has been 
studied. It is the aim of this paper to investigate the spectral flow of 
the Dirac Hamiltonian in the background of crossing vortices.

As space-time manifold we choose the four-torus $\T^4$. 
It can be seen as $\R^4$ 
modulo the lattice
$$
\L^4 = \left\{x \in \R^4 \, | \, x = \sum_\mu n_\mu e_\mu \, , \, 
n_\mu \in \Z \right\} \, ,
$$
where $\{ e_\mu \} \, , \, \mu = 1,\ldots ,4$ are four orthogonal
basis vectors with lengths $L_\mu = |e_\mu|$.

\begin{figure}
\begin{minipage}{7cm}
\centerline{\epsfxsize=7 cm\epsffile{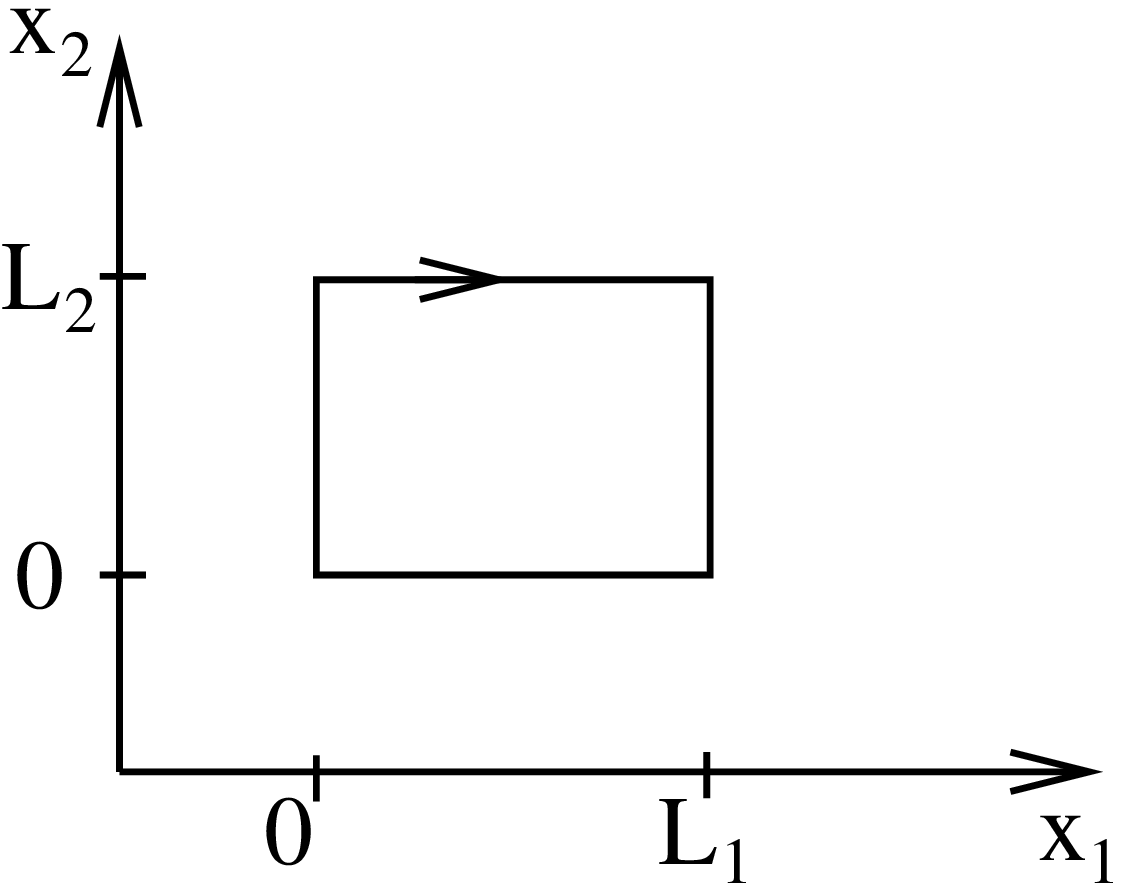}}
\caption{\label{adia-flux}
Wilson loop encircling the whole $x_1$-$x_2$-plane.}
\end{minipage}
\hfill
\begin{minipage}{8cm}
\centerline{\epsfxsize=8 cm\epsffile{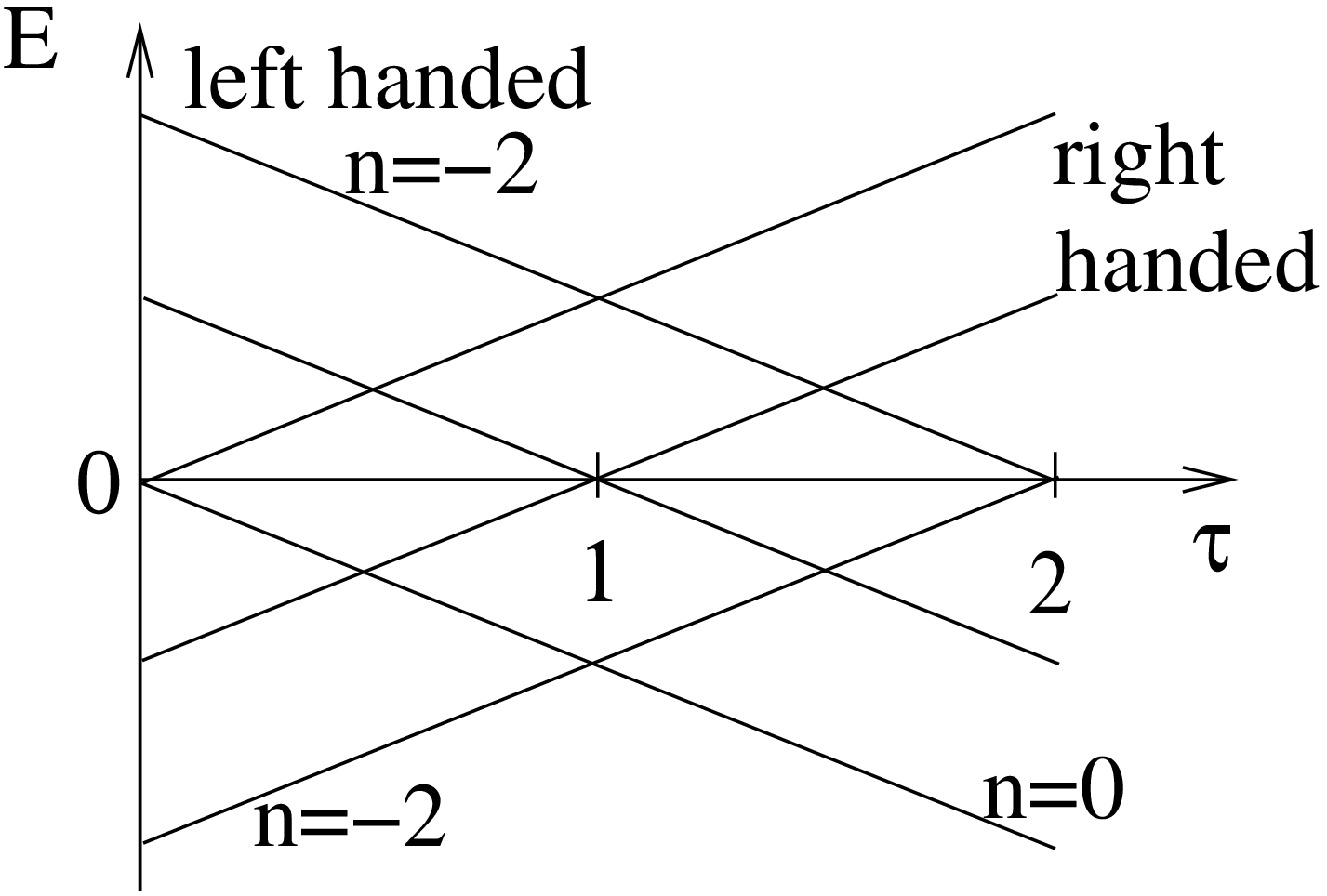}}
\caption{\label{spectralflow-const}
Illustration of the spectral flow of left and right handed fermions in 
topologically non-trivial constant background 
field strength as function of time $\tau$.}
\end{minipage}
\end{figure}

To demonstrate how the spectral flow emerges for intersecting center
vortices we start with the following $SU(2)$ gauge field configuration 
on the four-torus $\T^4$
\begin{eqnarray}
\label{gaugepot-1}
A_1 = - 2 \pi i x_2 / (L_1 L_2) \tau_3 \, , \quad 
A_2 = 0 \, , \quad
A_3 = - 2 \pi i x_4 / (L_3 L_4) \tau_3 \, , \quad 
A_4 = 0 \, .
\end{eqnarray}
This gauge potential describes two intersecting ``vortices'' whose
fluxes are spread out over the whole $x_1$-$x_2$- and $x_3$-$x_4$-,
respectively, planes. Indeed, this configuration has constant non-zero 
field strength components
\begin{equation}
\label{field-strength-const}
F_{1 2}=2 \pi i / (L_1 L_2) \tau_3 \, , \, 
F_{3 4}=2 \pi i / (L_3 L_4) \tau_3
\end{equation}
so that their fluxes are given by
\begin{eqnarray}
\int \d \Sigma_{12} F_{12} &=& 
\int_0^{L_1} \d x_1 A_1 (x_2 = L_2) = - i 2 \pi  \tau_3 \, , \\
\int \d \Sigma_{34} F_{34} &=& 
\int_0^{L_3} \d x_3 A_3 (x_4 = L_4) = - i 2 \pi  \tau_3 \, , \\
\end{eqnarray}
which is twice the flux of a center vortex. The line integrals appearing
here represent the only non-vanishing contributions from the Wilson
loops encircling the whole $x_1$-$x_2$- and $x_3$-$x_4$-plane,
respectively, see fig.~\ref{adia-flux}.
We choose the flux of these configurations to be twice the flux of a
center vortex since a single center vortex cannot be accommodated by the
torus with fermions in the fundamental representation of the gauge
group, see the discussion given below eq.~(\ref{cocycle-twist}).

Furthermore, the configuration (\ref{gaugepot-1}) carries topological 
charge $\nu = -2$ which is homogeneously distributed over space-time, 
i.e.~the intersection ``point'' of these two very fat vortices is smeared 
out over the whole space-time. The gauge potential (\ref{gaugepot-1}) 
satisfies the boundary conditions
\begin{eqnarray}
\label{trans-A}
A_\nu ( x + e_\mu ) &=& A_\nu^{U_\mu} (x) = 
U_\mu (x) A_\nu (x) U_\mu^{-1} (x) + 
U_\mu (x) \partial_\nu U_\mu^{-1} (x) \, , 
\end{eqnarray}
where the transition functions $U_\mu$ can be
chosen as follows:
\begin{equation}
\label{transition-1}
U_1 = U_3 = \Id \, , \quad 
U_2 = \exp{(2 \pi i x_1/L_1 \tau_3)} \, , \quad 
U_4 = \exp{(2 \pi i x_3/L_3 \tau_3)} \, .
\end{equation}
In general the transition functions have to fulfill the cocycle
condition
\begin{eqnarray}
\label{cocycle-twist}
U_\nu (x + e_\mu) U_\mu (x) Z_{\mu \nu} = 
U_\mu (x + e_\nu) U_\nu (x) \, ,
\end{eqnarray}
where $Z_{\mu \nu}$ are elements from the center of the gauge group.
Throughout this paper we consider fermions in the fundamental
representation of $SU(2)$. This implies that all $Z_{\mu \nu}$ are equal
to $\Id$, because fermions in the fundamental representation cannot 
tolerate twisted boundary conditions which imply the presence of a 
single center vortex. For later use we also note that under a gauge 
transformation $A_\mu \to A_\mu^V$ the transition functions transform 
\begin{equation}
\label{gaugetransf-trans}
U^V_\mu(x) = V(x + e_\mu) U_\mu(x) V(x)^{-1} \, . 
\end{equation}
Furthermore, since our gauge configuration (\ref{gaugepot-1}) satisfies
the Weyl gauge $A_4 = 0$ its topological charge (\ref{top-charge}) can
be expressed in terms of the (in general non-Abelian) Chern-Simons
action $S_{\mathrm CS} (x_4)$ as \cite{Reinhardt:2001kf}  
\begin{equation}
\label{top-charge2}
\nu = \int_0^{L_4} \d x_4 \partial_4 S_{\mathrm CS} (x_4) \, .
\end{equation}
For well localized center vortices (representing closed flux loops in
$D=3$) the Chern-Simons action is nothing but their writhing number
\cite{Reinhardt:2001kf}. For the particular gauge potential 
(\ref{gaugepot-1}) the Chern-Simons action reduces to
\begin{equation}
\label{CS-1}
S_{\mathrm CS} (x_4) = 
- \frac{1}{4 \pi^2} \int_{\T^3} \tr ( A_3 F_{12} ) \d^3 x =
- 2 \frac{x_4}{L_4} := - 2 \tau \, 
\end{equation}
so that eq.~(\ref{top-charge2}) yields indeed $\nu = -2$ when the vortex
evolves from $\tau = 0$ to $\tau=1$.


We are interested in the time-evolution of fermions in the intersecting
center vortex background described by the gauge potential 
(\ref{gaugepot-1}). Specifically we want to study the time-evolution 
of the spectrum of the three dimensional Dirac-Hamiltonian
$$
H ( \tau ) \psi = E (\tau) \psi \, ,
$$
i.e.~the Dirac energies $E(\tau)$ as function of the dimensionless 
time $\tau$. 

The Dirac Hamiltonian of a massless fermion 
reads\footnote{The representation of the Dirac matrices used in this
paper can be found in Appendix \ref{conventions}.}
\begin{equation}
\label{Dirac-Hamiltonian}
H = - \gamma_4 \gamma_i \D_i = \left( \begin{array}{cc}
                - i \sigma_i \D_i & 0 \\
                0 & i \sigma_i \D_i
                \end{array} \right) \, , 
\end{equation}
where $\D_i = \partial_i + A_i$ is the covariant derivative. 
Under gauge transformations $A \to A^V$ the Dirac spinors in the 
fundamental representation of $SU(2)$ transform as
\begin{equation}
\label{gaugetrf-spinor}
\psi(x) \to \psi^V ( x)  =  V (x) \psi ( x ) \, .
\end{equation}
In view of equation (\ref{trans-A}) this implies the following 
boundary conditions for fermionic fields 
\begin{equation}
\label{period-spinor}
\psi ( x+ e_\mu )  =  U_\mu (x) \psi ( x ) \, .
\end{equation}
Because the gauge potential (\ref{gaugepot-1}) and the transition 
functions (\ref{transition-1}) are purely Abelian the two color components 
of the fermionic field decouple from each other. Therefore it is 
sufficient to consider only 
one of the two color components, i.e.~we restrict ourselves to spinors 
with the lower color component equal zero. To get the desired results
for spinors with the opposite color one has to change the sign of the
energy and to exchange the upper with the lower color component. 
If not stated otherwise, in the following we will always consider the
upper color component of the Dirac field. 

Coming back to the Dirac 
spectrum one first observes that it is invariant under the shift 
$\tau \to \tau+1$. This is because the two corresponding gauge 
potentials are related by the periodic gauge transformation 
$V=\exp{(-2 \pi i x_3 / L_3 \tau_3)}$ which leaves the above 
introduced transition functions $U_{1 / 2 / 3}$ invariant, 
see eq.~(\ref{gaugetransf-trans}). 

Since $H$ commutes with $\gamma_5$ the eigenvectors of $H$ can be chosen
as eigenvectors of $\gamma_5$. Hence the eigenvectors of $H$ 
carry good chirality. It remains to find the eigenvectors 
$\bar \psi$ of the Dirac operator 
\begin{equation}
\label{Dirac-3}
\DD{3} = i \sigma_i \D_i \, , \, \DD{3} \bar \psi = E \bar \psi
\end{equation}
in three dimensions. Because of the periodicity in the 
$3$-direction ($U_3 = \Id$) one can make the ansatz
$$
\bar \psi_n (x_1,x_2,x_3) = 
\tilde \psi_n (x_1,x_2) \exp{(- 2 \pi i n x_3 / L_3)} \, .
$$
Inserting this ansatz into the eigenvalue equation (\ref{Dirac-3})
the latter reduces to the eigenvalue problem
\begin{eqnarray}
\label{D2'}
\left( \DD{2} + \bar E_n \sigma_3 \right) 
\tilde \psi_n = 
\lambda \tilde \psi_n \, , \, \bar E_n = 2 \pi \frac{\tau + n}{L_3} \, ,
\end{eqnarray}
where $\DD{2} = i \left( \sigma_1 \D_1 + \sigma_2 \D_2 \right)$ is the
massless Dirac operator in two dimensions.
For the gauge potential (\ref{gaugepot-1}) the eigenvectors and
eigenvalues of $\DD{2}$ are known \cite{Azakov:1997xk}:
\begin{equation}
\label{spec-D2-const}
\DD{2} \tilde \phi_m = \tilde E_m \tilde \phi_m \, , \, 
\tilde E_m = \sign(m) \sqrt{\frac{4 \pi m}{L_1 L_2}} \, , \, 
m \in \Z \, .
\end{equation} 
According to the index theorem for Abelian gauge fields in $D=2$ 
dimensions, (for each color component) there is exactly one zero 
mode of  $\DD{2}$, namely $\tilde \phi_{m=0}$. This zero mode is 
left handed, 
i.e.~only the lower component of this two-spinor is non-zero.
Therefore, the two-spinor $\tilde \psi_{n m=0} = 
\tilde \phi_{m=0} (x_1,x_2) \exp{(- 2 \pi i n x_3 / L_3)}$ 
is eigenvector of $\DD{3}$ with eigenvalue $- \bar E_n$,
see eq.~(\ref{D2'}). The explicit form of the function 
$\tilde \phi_{m=0}$ is written down in eq.~(\ref{phi-0}).
For $m \neq 0$  the eigenvectors $\tilde \psi_n$ (\ref{D2'}) are given
by linear combinations of $\tilde \phi_m$
and $\tilde \phi_{-m}$ 
$$
\tilde \psi_{n m} = \alpha_{n m} \tilde \phi_m + 
\beta_{n m} \tilde \phi_{-m}
$$
Putting this ansatz into eq.~(\ref{D2'}) and noting 
that\footnote{This is because  $\DD{2}$ and $\sigma_3$ anti-commute.} 
$\sigma_3 \tilde \phi_m = \phi_{-m}$ one gets the following 
eigenvalue problem for the vector $(\alpha_{n m},\beta_{n m})$:
\begin{eqnarray}
\label{2D-eigenvalue-problem}
\left(\begin{array}{cc}
\tilde E_m & \bar E_n \\
\bar E_n & - \tilde E_m
\end{array}
\right)
\left(\begin{array}{c}
\alpha_{n m} \\ 
\beta_{n m}
\end{array}
\right)
&=& E_{n m} 
\left(\begin{array}{c}
\alpha_{n m} \\ 
\beta_{n m}
\end{array}
\right) \, .
\end{eqnarray}
This eigenvalue problem has for fixed $(n,m)$ two solutions:
\begin{eqnarray}
\label{energy-1}
E_{n m} &=& \pm \sqrt{{\tilde E_m}^2 + {\bar E_n}^2} \, ,
\end{eqnarray}
where the sign is given by the sign of $\tilde E_m$.
One of the two solutions corresponds to positive $m \in \Z$ and the other 
to negative $m \in \Z$. Therefore we set $E_{n m} = \tilde E_m 
\sqrt{1 + (\bar E_n / \tilde E_m)^2} \, , \, m \neq 0$. The corresponding 
coefficients $\alpha_{n m}$ and $\beta_{n m}$ can be calculated from
(\ref{2D-eigenvalue-problem}).

Using the tensor product notation $\tilde \phi \otimes \chi_l$ 
for right handed ($l=1$) and left handed ($l=2$) four-spinors
\begin{equation}
\label{spinor-tensor}
\tilde \phi \otimes \chi_1 = 
\left(\begin{array}{c}
\tilde \phi \\
0 \\
0
\end{array}\right) \, , \quad
\tilde \phi \otimes \chi_2 = 
\left(\begin{array}{c}
0 \\
0 \\
\tilde \phi 
\end{array}\right) \, ,
\end{equation}
the eigenvalues and eigenfunctions of the Dirac Hamiltonian $H$ 
(\ref{Dirac-Hamiltonian}) can be written as
\begin{equation}
\label{spectral-flow1}
E_{m n l} = 
(-1)^{l} \left\{ \begin{array} {ll} 
\displaystyle
\tilde E_m 
\sqrt{ 1 + \frac{4 \pi^2 (n+\tau)^2}{L_3^2 {\tilde E_m}^2}} 
\, , \, & m \neq 0 \\
\displaystyle
- \frac{2 \pi}{L_3} (n+\tau) \, , \, & m = 0
\end{array} \right\} \, ,
\end{equation}
\begin{eqnarray}
\label{ev-H-const-1}
\psi_{n m l} &=& \left( \alpha_{n m} \tilde \phi_m 
+ \beta_{n m} \tilde \phi_{-m} \right) \otimes \chi_l \,
e^{(- 2 \pi i n x_3 / L_3)} 
\, , \, m,n \in \Z \, , \, m \neq 0 \, , \, l=1,2 \, , \\
\label{ev-H-const-2}
\psi_{n 0 l} &=& \tilde \phi_0 \otimes \chi_l \,
e^{(- 2 \pi i n x_3 / L_3)} 
\, , \, n \in \Z \, , \, l=1,2 \, . 
\end{eqnarray}
Obviously, only the $m=0$ level can cross zero energy. 
As the time $\tau$ evolves from $\tau=0$ to $\tau=1$ the left handed 
($l=2,m=0$) modes show a spectral flow from
positive to negative energy and the right handed ($l=1,m=0$) modes from
negative to positive energy, and there is a level crossing through zero 
energy for $n=-\tau$, see fig.~\ref{spectralflow-const}.
The eigenmodes of the lower color component show the opposite behavior.

Semi-classically, one can interpret the spectral flow induced by 
the time-evolution 
$\tau \to \tau+1$ as the creation of a (right handed) particle and a
(left handed) anti-particle, i.e.~the number of right-handed particles  
increases by $2$. Correspondingly the Chern-Simons action 
$S_{\mathrm CS} (\tau)$ changes by $-2$.

\begin{figure}
\begin{minipage}{10cm}
\centerline{\epsfxsize=9 cm\epsffile{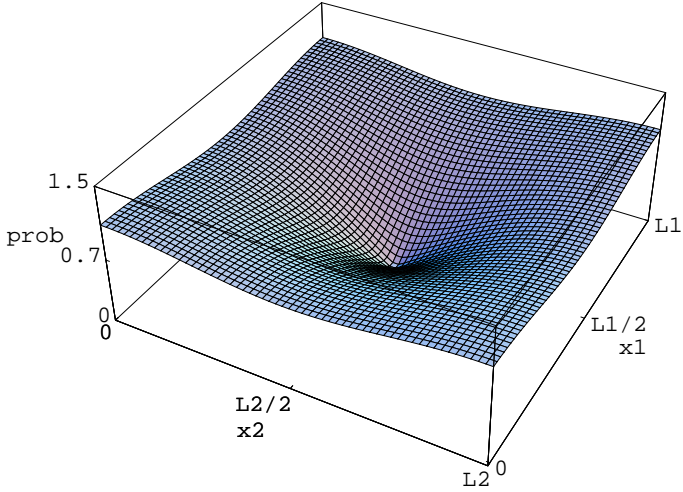}}
\caption{\label{prob-const}
Probability density of the mode $\psi_{002}$ at constant $x_3$.}
\end{minipage}
\end{figure}

The probability density of the spinor $\psi_{n=0 , m=0 , l=2}$ 
is plotted in fig.~\ref{prob-const} for fixed $x_3$. This density is 
independent of $x_3$ and of the adiabatic parameter $\tau$. 
The probability density has a zero and, obviously, breaks the translational 
invariance of the field strength\footnote{This phenomenon is already
known from the localized Landau orbits in a constant magnetic field.} 
(\ref{field-strength-const}). In 
\cite{Reinhardt:2002cm} it has been shown that the position of this zero
can be moved by adding a constant background gauge potential which does
not change the field strength and the boundary conditions.

Let us now come back to the Dirac operator on the four-torus. 
From the index theorem in $D=4$ 
it is  known that there is exactly one left handed zero mode $\psi_0$ 
of the Dirac operator with the above defined gauge potential. This zero 
mode can be explicitly written down \cite{Reinhardt:2002cm} and is given
in appendix \ref{zeromode-1}. 
One can show that this zero mode $\psi_0$ can be written as the sum over 
all left handed eigenmodes $\psi_{n m l}$ of the Dirac Hamiltonian 
$H$ which cross zero energy as the time $\tau$ passes from $\tau=-\infty$ 
to $\tau=+\infty$, which are the states with the quantum numbers 
$m=0 , l=2$. The time dependence 
of a solution $\tds_{n 0 l} = f_{n 0 l}(x_4) \psi_{n 0 l}$ of the 
(Euclidean) time-dependent Dirac equation
\begin{equation}
\partial_4 \tds_{n 0 l} = H \tds_{n 0 l} 
\end{equation}
is given by
\begin{eqnarray}
\partial_4 f_{n 0 l} (x_4) &=& E_{n 0 l} f_{n 0 l} (x_4) \, , \\
f_{n 0 l} (x_4) &=& 
C_{n 0 l} \exp{\left\{ 
- (-1)^{l} \pi \frac{L_4}{L_3} \left( n + \frac{x_4}{L_4}\right)^2
\right\} } \, .
\end{eqnarray}
The right-handed ($l=1$) solutions are exponentially growing 
for $x_4 \to \pm \infty$ whereas the left-handed solutions ($l=2$) are 
exponentially decreasing. Therefore one can construct a solution which 
is quasi-periodic\footnote{The solution will be periodic under 
$x_4 \to x_4+L_4$ up to a gauge transformation with the transition function 
$U_4$, see eq.~(\ref{transition-1}).} 
in $x_4$ from the left-handed solutions $\tds_{n 0 2}$ by 
a simple summation over all modes $n \in \Z$, putting 
$C_{n 0 2}=1 , \forall n \in \Z$:
\begin{eqnarray}
\nonumber
\psi (x) &=& \sum_{n \in \Z} \tds_{n 0 2} (x) =  
\tilde \phi_0 (x_1,x_2) \otimes \chi_2 \sum_{n \in \Z} 
\exp{\left\{ 
- \pi \frac{L_4}{L_3} \left( n + \frac{x_4}{L_4}\right)^2
 - 2 \pi i n \frac{x_3}{L_3} \right\} } \\
\nonumber
&=& 
\tilde \phi_0 (x_1,x_2) \otimes \chi_2 
\exp{\left\{- \frac{\pi x_4^2}{L_3 L_4}\right\} } 
\sum_{n \in \Z}
\exp{\left\{ - \pi \frac{L_4}{L_3} n^2 - 2 \pi i n 
\left( \frac{x_3}{L_3} - i \frac{x_4}{L_3} \right) \right\}}
\\
\label{summation}
&=&
\label{relation-zero-1}
\tilde \phi_0 (x_1,x_2) \otimes \chi_2 
\exp{\left\{- \frac{\pi x_4^2}{L_3 L_4}\right\} } 
\overline{\theta \left( \frac{x_3}{L_3} + i \frac{x_4}{L_3} , 
i \frac{L_4}{L_3} \right)} \, .
\end{eqnarray}
Here $\theta \left( z , i \tautheta \right)$ is the theta function which
is defined in eq.~(\ref{theta})
Comparing with eqs.~(\ref{phi-0},\ref{Dirac-zero-D4}) one observes 
that this is precisely the zero 
mode $\psi_0$ of the Dirac operator on the four-torus. In the limit 
$L_4/L_3 \gg 1$ and $|x_4 / L_4| \ll 1$ the part of the sum in 
eq.~(\ref{summation}) with $n=0$ dominates, i.e.~the sum can be 
approximated by $1$. Neglecting the $x_4$ dependence one obtains the 
eigenvector $\psi_{0 0 2}$ which crosses zero energy for 
$x_4=0$, i.e.~the zero mode on the four-torus approximates the 
time-evolution of that adiabatic eigenvector 
of $H$ who crosses zero energy during the evolution.
One reason for getting the strict 
relation (\ref{relation-zero-1}) between the zero mode on $\T^4$ and the
eigenmodes of the Dirac-Hamiltonian on $\T^3$ is that the eigenstates
$\psi_{n 0 2}$ are independent of $x_4$. Therefore the adiabatic 
approximation becomes exact and the $x_4$ dependent adiabatic solutions 
of the stationary Dirac equation are exact zero modes 
of the Dirac equation on $\T^3 \times \R$.

\mysection{Spectral flow for orthogonally intersecting vortices}
\label{orthogonal-thin}

In this chapter we consider two thin but smeared out 
vortices carrying both twice the flux of a center vortex in $SU(2)$.
We choose one vortex to be parallel to the $x_3$-$x_4$ plane and located 
at $x_1 = x_1^{(0)} , x_2 = x_2^{(0)}$ and the other parallel to the 
$x_1$-$x_2$ plane and located at $x_3 = x_3^{(0)} , x_4 = x_4^{(0)}$. 
The gauge potential of these two vortices can be written 
down in terms of theta functions \cite{Reinhardt:2002cm}. To this end
one introduces complex variables $u , v$:
\begin{eqnarray}
\begin{array}{rclcrcl}
\displaystyle
u &=& \left( x_1 + i x_2 \right) / L_1 & , & 
u^{(0)} &=& \left( x_1^{(0)} + i x_2^{(0)} \right) / L_1 \, , \\ 
\displaystyle
v &=& \left( x_3 + i x_4 \right) / L_3 
& , & 
v^{(0)} &=& \left( x_3^{(0)} + i x_4^{(0)} \right) / L_3 \, . 
\end{array}
\end{eqnarray} 
In terms of the new variable $u$
the gauge potential $A^{(1)}$ corresponding to the first vortex reads 
\cite{Reinhardt:2002cm}
\begin{eqnarray}
\label{gaugepot-orthogonal-12}
A^{(1)}_1 = -i \partial_2 \phi(u,u^{(0)},\tau_{12}) \, , \,
A^{(1)}_2 = i \partial_1 \phi(u,u^{(0)},\tau_{12}) \, , \, 
A^{(1)}_3 = A^{(1)}_4 = 0 \, ,
\end{eqnarray} 
where $\tau_{\mu \nu} = L_\nu / L_\mu$ and the profile function $\phi$ 
is defined in terms of theta functions\footnote{Our conventions 
concerning theta functions can be found in appendix \ref{conventions}.}:
\begin{eqnarray}
\nonumber
\phi (u,u^{(0)},\tau_{12}) &=& \frac{1}{4} 
\log{\left(
\theta^+ (u,u^{(0)},\tau_{12}) 
\overline{\theta^+(u,u^{(0)},\tau_{12})} + 
\theta^- (u,u^{(0)},\tau_{12}) 
\overline{\theta^-(u,u^{(0)},\tau_{12})}
\right)} \, , \\
\label{theta-0-pm}
\theta^\pm (u,u^{(0)},\tau_{12}) &=&  
\theta ( u + \ha + \iha \tau_{1 2} - u^{(0)} \pm \varepsilon 
, i \tau_{1 2}) 
\, , \quad 
\varepsilon \in \R_+ \, .
\end{eqnarray}
The parameter $\varepsilon$ is a measure for the thickness of the
vortex, $\varepsilon^2$ is approximately the ratio between the
transversal vortex extension and the area of the two torus. 
The gauge potential $A^{(2)}$ of the second vortex is analogously given
by
\begin{eqnarray}
\label{gaugepot-orthogonal-34}
A^{(2)}_3 = -i \partial_4 \phi(v,v^{(0)},\tau_{34}) \, , \,
A^{(2)}_4 = i \partial_3 \phi(v,v^{(0)},\tau_{34}) \, , \, 
A^{(2)}_1 = A^{(2)}_2 = 0 \, .
\end{eqnarray} 
Obviously the gauge potential of the complete vortex configuration is
given by the sum $A = A^{(1)} + A^{(2)}$. For the subsequent considerations 
it is important that the gauge potential of the first vortex is 
given by components $A_1$ and $A_2$ which are independent of
$x_{3},x_{4}$ and the gauge potential of the second one by 
components $A_3$ and $A_4$ which are independent of $x_{1},x_{2}$. 
This means that the covariant derivatives $\D_1$ and $\D_2$ depend only 
on $x_1,x_2$ and the covariant derivative $\D_3$ depends only 
on $x_3$ and $x_4$. Under this assumption the spectrum of the Dirac
operator $\DD{3} = i \sigma_k \D_k$ in $D=3$ dimensions can be written 
down if the spectrum of the Dirac operator $\DD{2}= i \left(\sigma_1
\D_1 + \sigma_2 \D_2 \right)$ on the two torus (spanned by the basis 
vectors $e_1,e_2$) is known. 

One further remark is in order here. In Euclidean space-time the 
Dirac Hamiltonian
(\ref{Dirac-Hamiltonian}) is usually calculated in the Weyl gauge 
$A_4^W=0$. It is possible to write down a gauge transformation $V$ 
which transforms the gauge potential given in 
eq.~(\ref{gaugepot-orthogonal-34}) into the Weyl gauge:
\begin{equation}
V(\vec x,x_4) = 
{\mathcal P} \exp \left( 
\int_{t=0}^{t=x_4} A_4 (\vec x,t) \d t
\right) \, , \, \vec x = (x_1,x_2,x_3) \, ,
\end{equation} 
where ${\mathcal P}$ denotes path ordering. The gauge transformation
$V$ will not change the transition functions $U_{1/2/3}$, 
because $A_4$ and therefore also $V$ are Abelian and 
periodic in these three directions, cf.~eq.~(\ref{gaugetransf-trans}). 
The gauge transformation $V$ will in general alter 
the transition function $U_4$. But the transformation of $U_4$ is irrelevant 
in this context, because we are only interested in the spectrum of the 
Dirac Hamiltonian at a constant time $x_4$. 
Therefore it makes no difference to calculate 
the eigenspinors and eigenvalues of $H$ in the original gauge given in 
eqs.~(\ref{gaugepot-orthogonal-12}-\ref{gaugepot-orthogonal-34}) or  
in the Weyl gauge - the difference is simply a  
gauge transformation which is periodic in the spatial directions.

\begin{figure}
\begin{minipage}{7cm}
\centerline{\epsfxsize=7 cm\epsffile{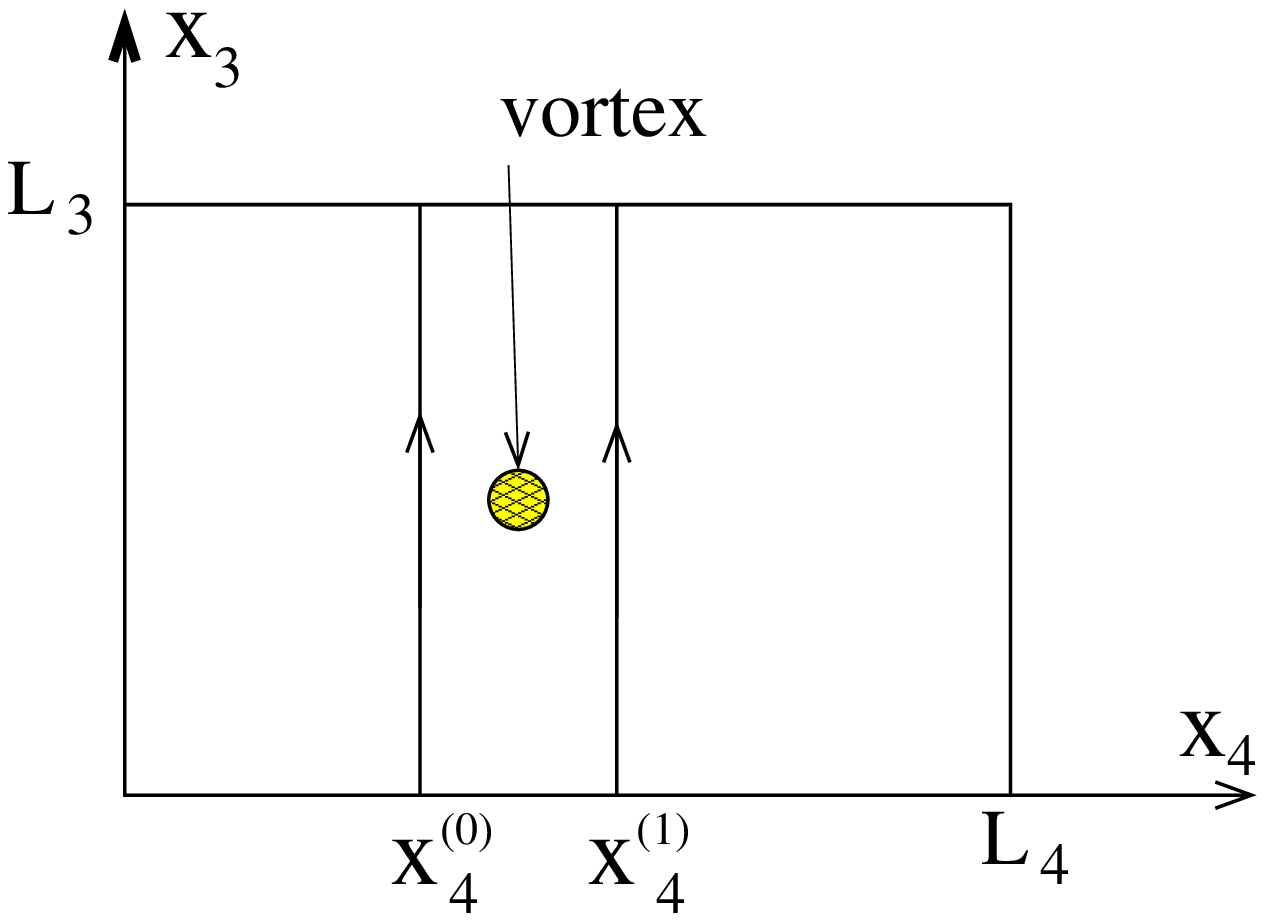}}
\caption{\label{fig2} Integration path on the two torus.}
\end{minipage}
\hfill
\begin{minipage}{7cm}
\centerline{\epsfxsize=7 cm\epsffile{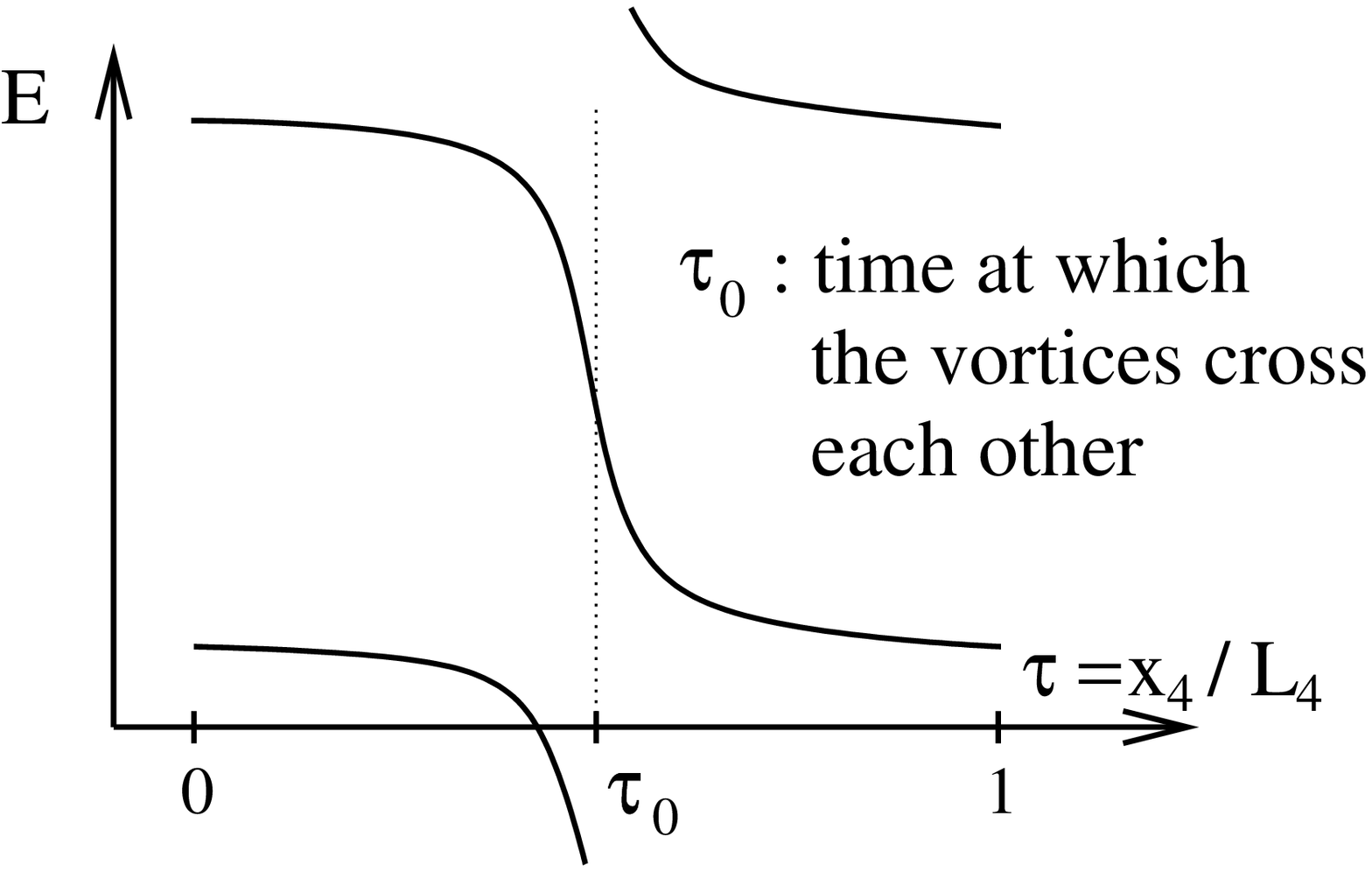}}
\caption{\label{adia-flux2} Spectral flow of the energy levels 
in the case of orthogonally intersecting vortices. Shown is the left
(right) handed energy level for the upper (lower) color component.}
\end{minipage}
\end{figure}

As in the case of constant field strength one can construct the eigenstates 
of $H$ with the help of the eigenstates 
$\{\tilde \phi_m \, , \, m \in \Z\}$ of the Dirac operator 
$\DD{2}=i\left(\sigma_1 \D_1 + \sigma_2 \D_2\right)$ 
with vortex background gauge potential on the two torus
$$
\DD{2} \phi_m = \tilde E_m \phi_m \, .
$$
As before, from the index theorem in $D=2$ dimensions it follows that there 
is exactly one left handed zero mode for the upper color index of $\DD{2}$ 
on the two torus, i.e.~$\tilde E_m=0$, if and only if $m=0$ 
(c.f.~eq.~(\ref{spec-D2-const})). Assuming that the eigenstates 
$\{\tilde \phi_m\}$ of $\DD{2}$ are known one can make the ansatz
(c.f.~eq.~(\ref{ev-H-const-1},\ref{ev-H-const-2}))
\begin{eqnarray}
\label{eigenvectors-1}
\psi_{n m l} &=& \left( \alpha_{n m} \tilde \phi_m
+ \beta_{n m} \tilde \phi_{-m} \right) \otimes \chi_l
\, \ff_n 
\, , \, m,n \in \Z \, , \, m \neq 0 \, , \, l=1,2 \, , \\
\label{eigenvectors-2}
\psi_{n 0 l} &=& \alpha_{n 0} \tilde \phi_0 \otimes \chi_l
\, \ff_n 
\, , \, n \in \Z \, , \, l=1,2 \, ,
\end{eqnarray}
where $\ff_n$ is an eigenfunction of $i \D_3$:
$$
i \D_3 \ff_n = \bar E_n \ff_n 
$$
which depends on $x_3$ and $x_4$.
Noting the periodicity of $\ff_n$ (because $U_3 = \Id$) one obtains the
solution
\begin{eqnarray}
\ff_n ( x_3 , x_4) &=&
\exp{ \left\{ - 2 \pi i n \frac{x_3}{L_3} 
+ \frac{x_3}{L_3} \int_0^{L_3} \d x_3' A_3 (x_3' , x_4 ) 
- \int_0^{x_3} \d x_3' A_3 (x_3' , x_4 ) \right\} } \, , \\
\bar E_n &=& 
\frac{2 \pi n}{L_3} + \frac{i}{L_3} \int_0^{L_3} \d x_3' A_3 (x_3')
\, .
\end{eqnarray}
Inserting this ansatz into the stationary Dirac equation 
$$
H \psi_{n m l} = E_{n m l} \psi_{n m l}
$$
yields (as in the case of constant field strength considered before) 
the eigenvalue equation (\ref{2D-eigenvalue-problem}) for the 
eigenvectors $(\alpha_{n m},\beta_{n m})$ and the energies $E_{n m l}$.
The energy eigenvalues are given by
\begin{eqnarray}
\label{spectral-flow2}
E_{n m l} &=& 
(-1)^l \left\{ \begin{array} {ll} 
\displaystyle
\tilde E_m \sqrt{1 + (b/ \tilde E_m)^2 } \, , \, & m \neq 0 \\
\displaystyle
- b \, , \, & m = 0
\end{array} \right\} \, , \\
\label{spectral-flow2-b}
b &=& 2 \pi \frac{n}{L_3} + \frac{i}{L_3} \int_0^{L_3} \d x_3 A_3 \, .
\end{eqnarray}
Obviously, only eigenvectors with $m=0$ can cross 
zero energy. The integral in eq.~(\ref{spectral-flow2-b}) is directly 
related to the field strength component $F_{34}$:
\begin{eqnarray}
\int_0^{L_3} \d x_3' A_3 (x_3',x_4^{(1)} ) &=& 
\int_0^{L_3} \d x_3' A_3 (x_3',x_4^{(0)} ) + 
\int_0^{L_3} \d x_3' \int_{x_4^{(0)}}^{x_4^{(1)}} \d x_4 '
F_{4 3} (x_3',x_4' ) \, .
\end{eqnarray}
Therefore, one obtains
\begin{eqnarray}
\label{spec-flow-grad}
E_{n 0 1} (x_4^{(1)}) &=& E_{n 0 1} (x_4^{(0)}) - (-1)^l 
\frac{i}{L_3} 
\int_0^{L_3} \d x_3' \int_{x_4^{(0)}}^{x_4^{(1)}} \d x_4 '
F_{4 3} (x_3',x_4' ) \, ,
\end{eqnarray}
i.e.~the energy $E_{n 0 l}(x_4)$ changes most rapidly where the field 
strength is peaked and this is at the location of the vortex in the 
$x_3$-$x_4$ plane (which is of course the time where the two orthogonal
vortices intersect on $\T^4$), see figs.~\ref{fig2} and 
\ref{adia-flux2}.
It turns out that left handed ($l=2$) and right handed ($l=1$) spinors 
cross zero energy level starting from positive to negative energies and 
vice versa. The time $x_4^{(0)}$ for which the 
crossing through zero energy takes place can be shifted if one adds a 
constant to the gauge potential $A_3$. This addition of a constant will 
not change the field strength, but it can obviously change the time at
which the zero energy crossing occurs. Therefore, the precise time 
$x_4^{(0)}$ of the level crossing seems to be irrelevant.

Obviously, in the case of crossing vortices the adiabatic approximation is 
(especially in the neighborhood of the vortex) not very good, because the
eigenstates $\psi_{n m l}$ depend strongly on $x_4$. Therefore in this case
we do not have such a simple relation between the zero mode on the four-torus
and the adiabatic eigenstates of the Dirac Hamiltonian, 
see eq.~(\ref{relation-zero-1}).

\begin{figure}
\begin{minipage}{7cm}
\centerline{\epsfxsize=7 cm\epsffile{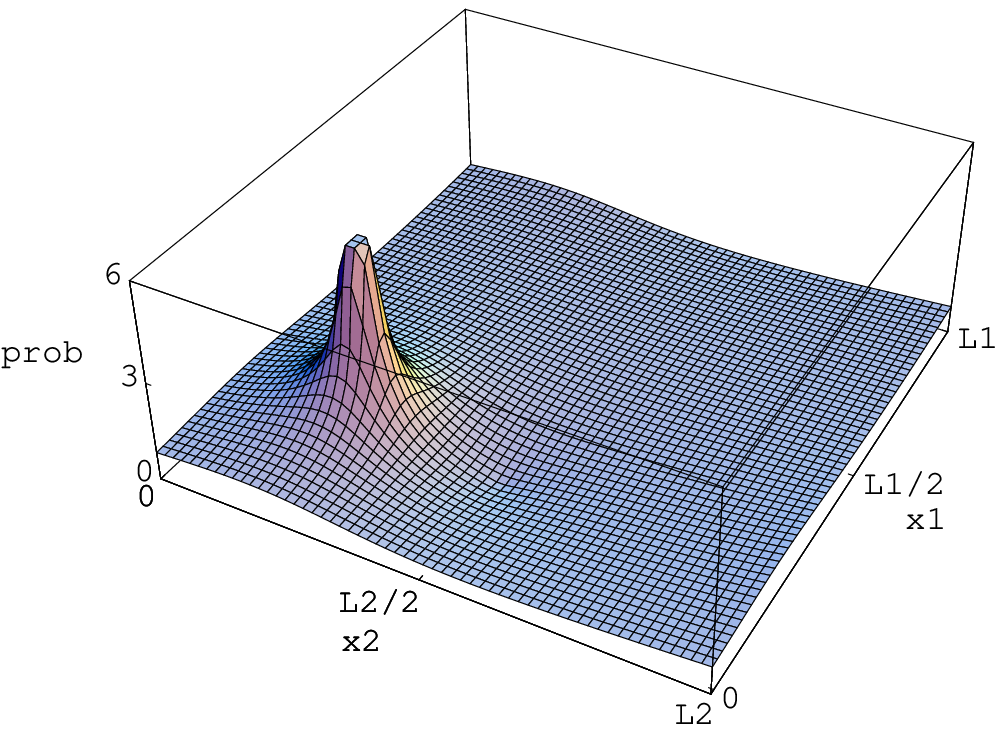}}
\caption{\label{prob-vortex} Probability density of the eigenvector 
$\psi_{002}$ at constant $x_3$.}
\end{minipage}
\hfill
\begin{minipage}{7cm}
\centerline{\epsfxsize=7 cm\epsffile{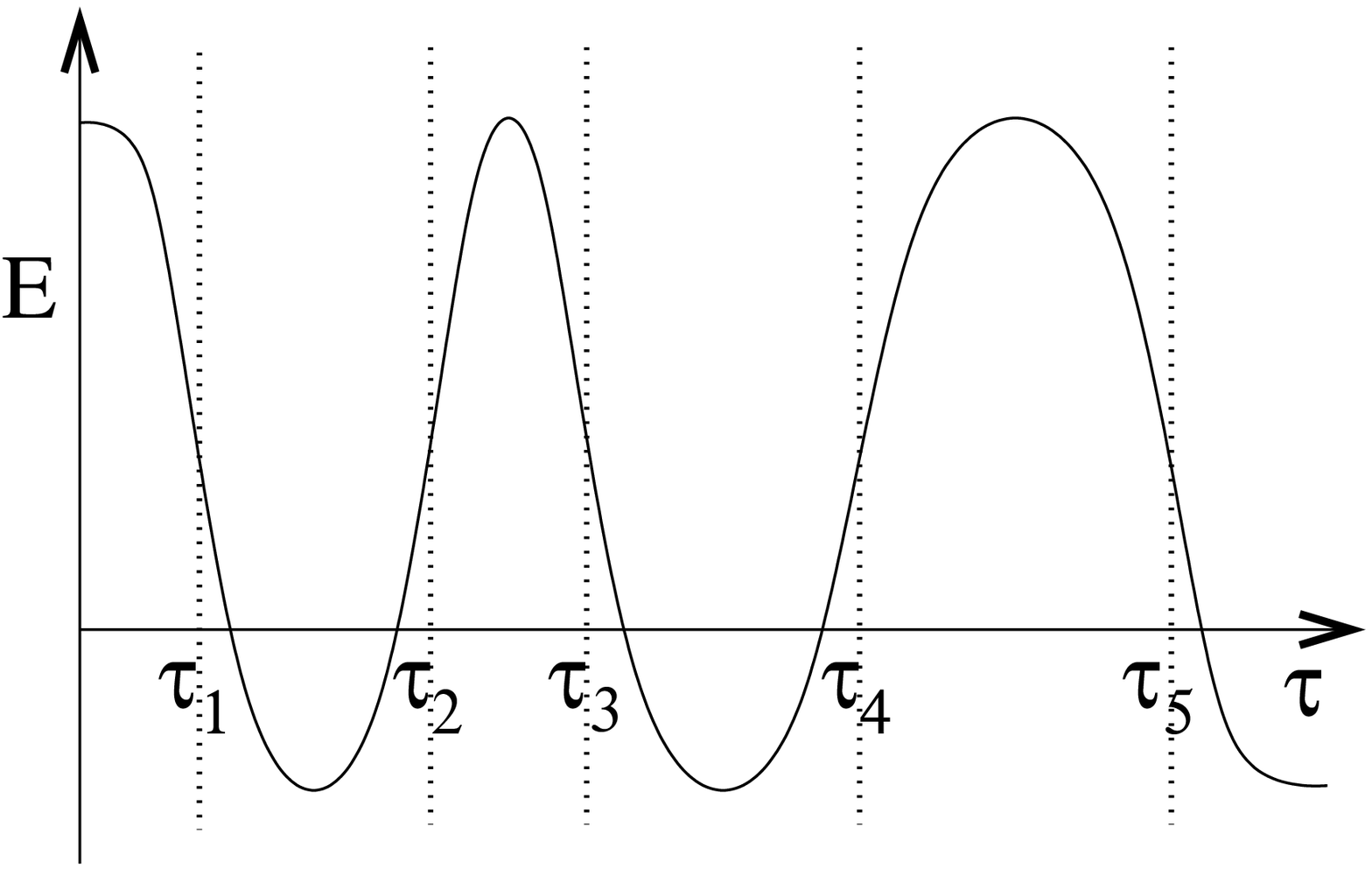}}
\caption{\label{spec-flow-gas} Spectral flow for $M=2$ vortices
and $N=3$ anti-vortices respectively located at times 
$\tau_k = x_4^{(k)} / L_4$.}
\end{minipage}
\end{figure}

The {\em probability density} of the eigenvectors $\psi_{nm2}$ is 
independent of $x_3$ and of the adiabatic parameter $x_4$. 
In fig.~\ref{prob-vortex} the probability density of the mode 
$\psi_{002}$ has been plotted for constant $x_3$. As in the case of
constant field strength there is a zero of the density. But, in
addition, here the density is peaked 
at the position of the vortex in the $x_1-x_2$ plane. 

It is easy to generalize the previous results to background
fields consisting of more than two intersecting vortices. 
In what follows we consider one vortex parallel to the $x_3$-$x_4$ plane 
and located at $x_1 = x_1^{(0)} , x_2 = x_2^{(0)}$ and a number $M$ of
vortices and  $N$ anti-vortices which are all parallel to the 
$x_1$-$x_2$ plane, located at $x_3 = x_3^{(k)} , x_4 = x_4^{(k)} , k
= 1 , \ldots , (M+N)$, and carrying magnetic flux $\phi_k = \pm 2 \pi$.
The topological charge of this configuration is given by the sum 
$\nu = \frac{1}{2 \pi} \sum_k \phi_k = M-N$ over the magnetic fluxes of 
the individual vortices piercing the $x_3$-$x_4$ plane. 
The gauge potential corresponding to these vortices is simply given by
the sum over the gauge potentials of the individual vortices, see
eq.~(\ref{gaugepot-orthogonal-34}). The eigenfunctions and eigenvalues 
of the Dirac Hamiltonian are given by 
eqs.~(\ref{eigenvectors-1}-\ref{spectral-flow2-b}) with gauge potential
$A_3$ replaced by the corresponding expression of the multiple vortex
configuration. The spectral flow is shown in fig.~\ref{spec-flow-gas} 
for $M=2$ vortices and $N=3$ anti-vortices. Obviously, the gradient 
of the spectral flow is maximal at times $x_4^{(k)}$ where two vortices 
cross each other. The sign of the gradient depends on the flux of the 
particular vortex located
at the time $x_4^{(k)}$, cf.~eq.~(\ref{spec-flow-grad}). 
For a gas of orthogonally intersecting vortices the spectrum of the
Dirac Hamiltonian fluctuates with time $\tau$, 
cf.~eq.~(\ref{spectral-flow2}). For $m=0$ ($\tilde E_m=0$) 
the energy eigenvalue fluctuates around zero.

Let us now see how these fluctuations in the spectral flow manifest
themselves in the spectrum of the $D=4$ Dirac operator. Considering a
multi-vortex configuration as given above with $M=N$, i.e.~with
topological charge $\nu=0$. A numerical investigation shows that the
module of the smallest eigenvalue of the $D=4$ Dirac operator decreases
with increasing number of vortex-anti-vortex pairs. It seams that the
fluctuations of the energy eigenvalues of the Dirac Hamiltonian  lower
the eigenvalue of the Dirac operator with smallest absolute value. It
remains to be seen whether this observation can provide a mechanism to
generate a non-zero level density at zero virtuality which is sufficient
for spontaneous breaking of chiral symmetry. This will be subject to
further research.

\mysection{Spectral flow for non-orthogonally intersecting vortices}
\label{non-orthogonal-thin}

\begin{figure}
\begin{minipage}{7cm}
\centerline{\epsfxsize=7 cm\epsffile{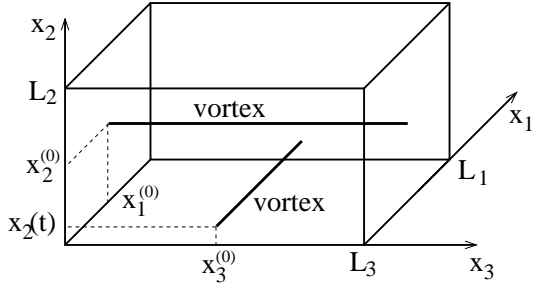}}
\caption{\label{fig1}
Location of the two vortices on the three-torus.}
\end{minipage}
\end{figure}

\begin{figure}
\begin{minipage}{15cm}
\centerline{\epsfig{file=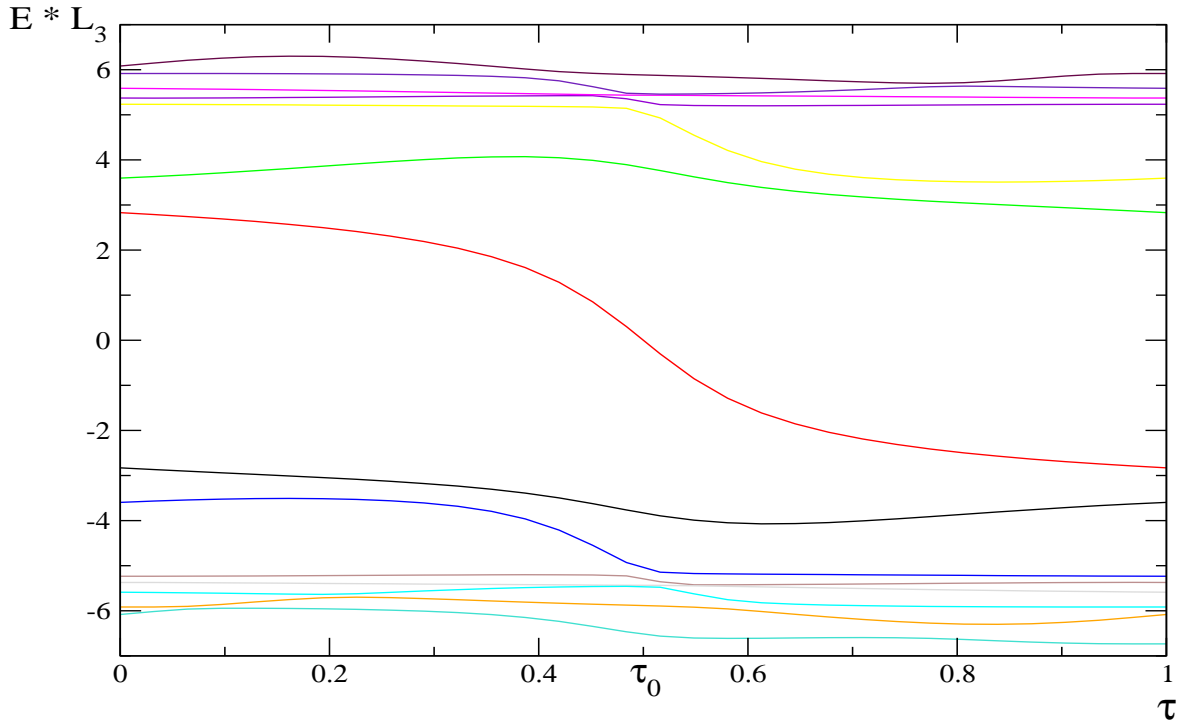,width=12cm,height=18cm,angle=-90}}
\caption{\label{spectral_flow}
Spectral flow of the left handed eigenmodes of 
the Dirac Hamiltonian for non orthogonally intersecting vortices
depending on $x_4/L_4$. The vortices cross at $x_4=L_4 / 2$,
i.e.~at $\tau=\tau_0=0.5$.}
\end{minipage}
\end{figure}

\begin{figure}
\begin{minipage}{7cm}
\centerline{\epsfxsize=7 cm\epsffile{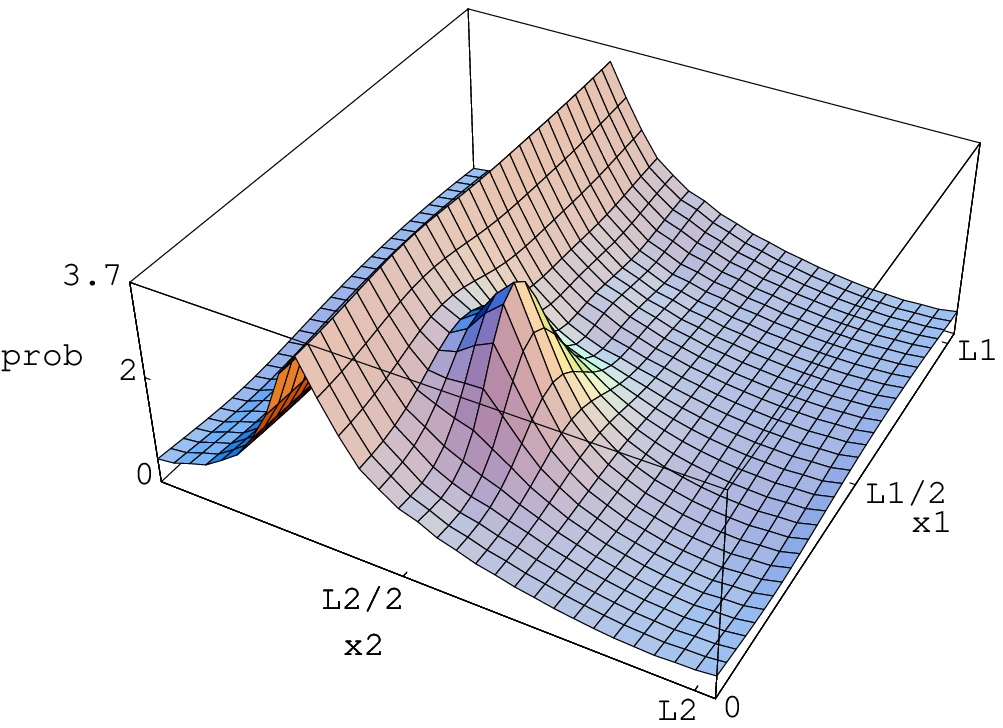}}
\end{minipage}
\hfill
\begin{minipage}{7cm}
\centerline{\epsfxsize=7 cm\epsffile{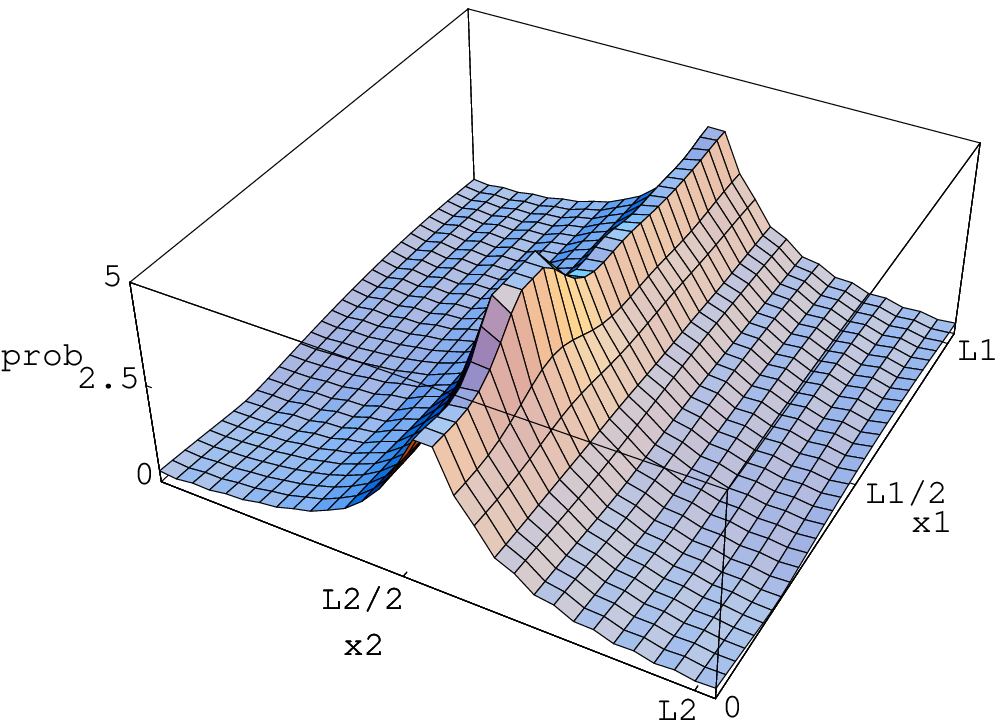}}
\end{minipage}
\caption{\label{zeromode2}
Probability density of the eigenmode of the Dirac Hamiltonian whose
energy eigenvalue crosses zero where the parameters in eq.~(\ref{param}) 
have been chosen to be $u^{(0)} = 0.3 + i 0.5$, $v^{(0)} = 0.3$ at 
$x_3 =0.3 L_3 , x_4 = 0.29$ and $x_3 =0.3 , x_4 = 0.52$, respectively.}
\end{figure}

As before we will consider a configuration on $\T^4$ which describes 
two intersecting center vortices. On a spatial torus $\T^3$ one vortex 
is static and parallel to the $x_3$-axis located at 
$x_1=x_1^{(0)}, x_2=x_2^{(0)}$. The other vortex is time-dependent and 
parallel to the $x_1$-axis and located at 
$x_3=x_3^{(0)},x_2=(L_2/L_4) x_4$, 
see fig.~\ref{fig1}. The two vortices intersect in $\T^4$ at 
$x_\mu=x_\mu^{(0)} , \mu = 1,\ldots,3 , x_4 = (L_4/L_2) x_2^{(0)}$. 
The gauge potential of the second vortex can be constructed from the
potential of a vortex which is originally parallel to the $x_1-x_2$ plane
by rotation in the $x_2$-$x_4$-plane. 
Note that the rotated vortex has magnetic
and electric field strength (the original vortex only has an electric
field strength).
To write down the gauge potential of the vortex configuration one 
again introduces complex variables $u , v$:
\begin{eqnarray}
\label{param}
\begin{array}{rclcrcl}
\displaystyle
u &=& \left( x_1 + i x_2 \right) / L_1 & , & 
u^{(0)} &=& \left( x_1^{(0)} + i x_2^{(0)} \right) / L_1 \, , \\ 
\displaystyle
v &=& \left( x_3 + i \left( x_4 - x_2 \tau_{24} \right) \right) / L_3 
& , & 
v^{(0)} &=& x_3^{(0)} / L_3 \, . 
\end{array}
\end{eqnarray} 
The gauge potential $A^{(1)}$ of the first vortex is again  given by
\cite{Reinhardt:2002cm}
\begin{eqnarray}
A^{(1)}_1 = -i \partial_2 \phi(u,u^{(0)},\tau_{12}) \, , \,
A^{(1)}_2 = i \partial_1 \phi(u,u^{(0)},\tau_{12}) \, , \, 
A^{(1)}_3 = A^{(1)}_4 = 0 \, .
\end{eqnarray} 
The gauge potential $A^{(2)}$ of the second (time-dependent) vortex 
is slightly more involved: 
\begin{eqnarray}
A_1 = 0 \, , \, 
A_2 = - i \tau_{24} \partial_3 \phi(v,v^{(0)},\tau_{34})  \, , \,
A_3 = - i \partial_4 \phi(v,v^{(0)},\tau_{34}) \, , \, 
A_4 = i \partial_3 \phi(v,v^{(0)},\tau_{34}) \, .
\end{eqnarray} 
The gauge field configuration $A=A^{(1)}+A^{(2)}$ has topological charge 
$\nu=1$ and the transition functions are given by
\begin{eqnarray}
U_1 = U_3 = \Id \, , \quad 
U_2 = \exp{\left( 2 \pi i  
      \left( \frac{x_1}{L_1} + \frac{x_3}{L_3} \right) \right)} 
\, , \quad 
U_4 = \exp{\left( 2 \pi i  \frac{x_3}{L_3} \right)} \, .
\end{eqnarray}
In this case the eigenvectors of the Dirac Hamiltonian $H$ do not
factorize as in the examples considered before. In a numerical study we
obtained a spectral flow of left handed four-spinors from positive to
negative energy in accordance with the existence of a left-handed zero
mode of the Dirac operator in $D=4$, see fig.~\ref{spectral_flow}. 
From the figure \ref{spectral_flow} one can argue that (as in the case 
of orthogonally intersecting vortices) the largest
gradient of the spectral flow with respect to time $\tau$ is where the
vortices intersect, $\tau = \tau_0$, i.e.~where the topological charge 
is localized. 
Further it seems that all eigenmodes cross zero energy level as time 
$\tau$ evolves from $-\infty$ to $+\infty$. This is different from the
case of orthogonally intersecting vortices considered before. There one
has different branches in the spectral flow, 
compare eq.~(\ref{spectral-flow2}),
and only eigenmodes with $m=0$ cross zero energy as time $\tau$ evolves 
from $-\infty$ to $+\infty$.
The probability densities of the eigenvectors of the Dirac Hamiltonian
again show an enhancement at the position of the vortices, see
fig.~\ref{zeromode2}. 

\mysection{Concluding remarks}
\label{conclusion}

We have studied the quark spectrum in the background of intersecting
flat vortex sheets. Choosing the four-torus $\T^4$ as space-time 
manifold we have calculated the spectral flow of the Dirac Hamiltonian as
function of time for intersecting vortex fields. 
The spectral flow as function of time $\tau$ has the
largest gradient in the neighborhood of the vortex intersection point,
i.e.~where the topological charge is localized. For the infinitely
thick intersecting ``vortices'' considered in section 
\ref{orthogonal-fat} this gradient is constant 
(see eq.~(\ref{spectral-flow1})) reflecting the fact that in this case 
the topological charge is homogeneously distributed over space-time.
These results are consistent with the findings of 
ref.~\cite{Reinhardt:2002cm} that zero modes of the 4-dimensional Dirac
operator are localized at the intersection points  of vortices.
Also, the probability density of the quark eigenmodes of the Dirac
Hamiltonian considered in the present paper show a strong correlation 
to the position of the vortex sheets. 
Depending on the energy level considered, the vortices seem to attract or 
repel the quarks. For the lowest lying states the probability density
always shows an enhancement at the position of the vortices, i.e.~the
vortices ``act attractively'' on the quarks in these modes.

\section{Acknowledgments}
 
The authors are grateful to O.~Schr\"oder for helpful discussions. 
This work has been supported by the Deutsche Forschungsgemeinschaft 
under grants DFG-Re 856/5-1.

\appendix

\mysection{Conventions}
\label{conventions}

We choose the generators of the gauge group to be anti-hermitian.
Therefore the components $A_\mu$ of the gauge potential are
anti-hermitian, e.g.~purely imaginary for the gauge group $U(1)$. The
magnetic flux $\Phi$ through a closed loop ${\mathcal C}$ is 
defined by 
\begin{equation}
\label{flux-def}
\Phi = \frac{1}{2 \pi i} \oint_{\mathcal C} A_\mu \d x_\mu 
\end{equation}
and thus real valued.

We consider the Dirac equation in Euclidean space-time. 
In $D=4$ we use the chiral representation for the Dirac matrices:
\begin{eqnarray}
\label{dirac-matrices-D4}
\gamma_i &=& \left( \begin{array}{cc}
                0 & \sigma_i \\
                \sigma_i & 0 
                \end{array} \right) \, , \quad i = 1,2,3 \, , \quad 
\gamma_4 = \left( \begin{array}{cc}
                0 & i \Id \\
                - i \Id & 0 
                \end{array} \right) \, , \\
\gamma_5 &=& \gamma_1 \gamma_2 \gamma_3 \gamma_4 = 
            \left( \begin{array}{cc}
                \Id & 0 \\
                0 & - \Id 
                \end{array} \right) \, , 
\end{eqnarray}
where $\Id$ is the $2 \times 2$ unit matrix and $\sigma_i$ are the
(hermitian) Pauli matrices. 

The massless Dirac Hamiltonian $H$ in three dimensions reads
\begin{equation}
H = - \gamma_4 \gamma_i \D_i = \left( \begin{array}{cc}
                - i \sigma_i \D_i & 0 \\
                0 & i \sigma_i \D_i
                \end{array} \right) \, ,
\end{equation}
where $\D_i = \partial_i + A_i$ is the covariant derivative. 

In this paper we use the theta function conventions introduced in
\cite{tata}:
\begin{equation}
\label{theta}
\theta ( z , i \tautheta ) = \sum_{n \in \Z} 
e^{- \pi \tautheta n^2 + 2 \pi i n z} \, , \quad z \in \C \, , \, 
\tautheta \in \R_+ \, .
\end{equation}
These functions are analytic in $z$ and obey the periodicity properties
\begin{equation}
\label{period-theta}
\theta ( z + 1 , i \tautheta ) = \theta ( z , i \tautheta ) \, , \quad 
\theta ( z + i \tautheta , i \tautheta ) = e^{\pi \tautheta - 2 \pi i z}
\theta ( z , i \tautheta ) \, .
\end{equation}
The only zeros of this function are at the points \cite{tata}
\begin{equation}
z = (m + 1/2) + (n + 1/2) i \tautheta \, , \quad m,n \in \Z \, .
\end{equation}

\mysection{Zero mode of the Dirac operator in $D=4$}
\label{zeromode-1}

We consider the gauge potential 
\begin{eqnarray}
A_1 &=& - 2 \pi i x_2 / (L_1 L_2)  \, , \quad 
A_2 = 0 \, , \\
A_3 &=& - 2 \pi i x_4 / (L_3 L_4) \, \, , \quad 
A_4 = 0 \, 
\end{eqnarray}
on the four-torus with transition functions
\begin{equation}
\label{transition}
U_1 = U_3 = \Id \, , \quad U_2 = \exp{(-2 \pi i x_1/L_1)} \, , \quad 
U_4 = \exp{(- 2 \pi i x_3/L_3)} \, .
\end{equation}
The topological charge of this configuration is $1$, i.e.~there is one
(left handed) zero mode of the corresponding Dirac operator which can be
written down explicitly in terms of theta functions. 
To this end one introduces complex variables 
$u := ( x_1 + i x_2 ) / L_1 \, , \, v := (x_3 + i x_4 )/L_3$ and positive
real numbers $\tau_{12} := L_2 / L_1 \, , \, \tau_{34} := L_4 / L_3$. 
The problem of finding the zero mode becomes a problem of finding the
zero mode of the Dirac operator 
$\DD{2} = i \sigma_1 D_1 + i \sigma_2 D_2$  on the two-torus and this problem
can be easily solved (compare \cite{Reinhardt:2002cm,Azakov:1997xk}). 
The zero mode of $\DD{2}$ is given by 
\begin{equation}
\label{phi-0}
\phi_0 = \left( \begin{array}{c}
                 0 \\
		 \zero(x_1,x_2,L_1,L_2)
		 \end{array} \right) \, , \quad 
\zero(x_1,x_2,L_1,L_2) = 
\exp{\left(-\frac{\pi x_2^2}{L_1 L_2}\right)} 
\overline{\theta(u , i L_2/L_1)} \, .
\end{equation} 
A short calculation shows that the zero mode in $D=4$ dimensions has the 
form:
\begin{equation}
\label{Dirac-zero-D4}
\psi_0 = \left( \begin{array}{c}
                 0 \\
		 0 \\
		 0 \\
		 \zero(x_1,x_2,L_1,L_2) 
		 \zero(x_3,x_4,L_3,L_4)
		 \end{array} \right) \, .
\end{equation} 
Obviously, the non-zero spinor component of $\psi_0$ is the product of 
the two zero mode functions $\zero$ from $D=2$.


\begin{thebibliography}{10}

\bibitem{'tHooft:1978hy}
G.~'t~Hooft.
\newblock On the phase transition towards permanent quark confinement.
\newblock {\em Nucl. Phys.}, B138:1, 1978.

\bibitem{Mack:1979rq}
G.~Mack and V.~B.~Petkova.
\newblock Comparison of lattice gauge theories with gauge groups Z(2) and
  SU(2).
\newblock {\em Ann. Phys.}, 123:442, 1979.

\bibitem{Greensite:2003bk}
J.~Greensite.
\newblock The confinement problem in lattice gauge theory.
\newblock {\em hep-lat/0301023}.

\bibitem{Engelhardt:1999fd}
M.~Engelhardt, K.~Langfeld, H.~Reinhardt, and O.~Tennert.
\newblock Deconfinement in SU(2) Yang-Mills theory as a center vortex
  percolation transition.
\newblock {\em Phys. Rev.}, D61:054504, 2000.

\bibitem{Kovacs:2000sy}
T.~G.~Kovacs and E.~T.~Tomboulis.
\newblock Computation of the vortex free energy in SU(2) gauge theory.
\newblock {\em Phys. Rev. Lett.}, 85:704--707, 2000.

\bibitem{DelDebbio:1997mh}
L.~del~Debbio, M.~Faber, J.~Greensite, and S.~Olejnik.
\newblock Center dominance and Z(2) vortices in SU(2) lattice gauge theory.
\newblock {\em Phys. Rev.}, D55:2298--2306, 1997.

\bibitem{deForcrand:1999ms}
P.~de~Forcrand and M.~D'Elia.
\newblock On the relevance of center vortices to QCD.
\newblock {\em Phys. Rev. Lett.}, 82:4582--4585, 1999.

\bibitem{Karsch:1998ua}
F.~Karsch.
\newblock Deconfinement and chiral symmetry restoration.
\newblock {\em hep-lat/9903031}

\bibitem{Gattringer:2002dv}
C.~Gattringer, P.~E.~L.~Rakow, A.~Sch\"afer, and W.~Soldner.
\newblock Chiral symmetry restoration and the Z(3) sectors of QCD.
\newblock {\em Phys. Rev.}, D66:054502, 2002.

\bibitem{Engelhardt:1999wr}
M.~Engelhardt and H.~Reinhardt.
\newblock Center vortex model for the infrared sector of Yang-Mills theory:
  Confinement and deconfinement.
\newblock {\em Nucl. Phys.}, B585:591--613, 2000.

\bibitem{Engelhardt:2001ze}
M.~Engelhardt, M.~Faber, and H.~Reinhardt.
\newblock Center vortex model for the infrared sector of Yang-Mills theory.
\newblock {\em Nucl. Phys. Proc. Suppl.}, 106:655--657, 2002.

\bibitem{Reinhardt:2001kf}
H.~Reinhardt.
\newblock Topology of center vortices.
\newblock {\em Nucl. Phys.}, B628:133--166, 2002.

\bibitem{Engelhardt:1999xw}
M.~Engelhardt and H.~Reinhardt.
\newblock Center projection vortices in continuum Yang-Mills theory.
\newblock {\em Nucl. Phys.}, B567:249, 2000.

\bibitem{Cornwall:1999xw}
J.~M.~Cornwall.
\newblock Center vortices, nexuses, and fractional topological charge.
\newblock {\em Phys. Rev.}, D61:085012, 2000.

\bibitem{Atiyah:1980jh}
M.~F.~Atiyah, V.~K.~Patodi, and I.~M.~Singer.
\newblock Spectral asymmetry and Riemannian geometry 3.
\newblock {\em Math. Proc. Cambridge Phil. Soc.}, 79:71, 1980.

\bibitem{Reinhardt:2002cm}
H.~Reinhardt, O.~Schr\"oder, T.~Tok, and V.~C.~Zhukovsky.
\newblock Quark zero modes in intersecting center vortex gauge fields.
\newblock {\em Phys. Rev.}, D66:085004, 2002.

\bibitem{Christ:1980zm}
N.~H.~Christ.
\newblock Conservation law violation at high-energy by anomalies.
\newblock {\em Phys. Rev.}, D21:1591, 1980.

\bibitem{Khoze:1995yb}
V.~V.~Khoze.
\newblock Fermion number violation in the background of a gauge field in
  Minkowski space.
\newblock {\em Nucl. Phys.}, B445:270--294, 1995.

\bibitem{Klinkhamer:2001cp}
F.~R.~Klinkhamer and Y.~J.~Lee.
\newblock Spectral flow of chiral fermions in nondissipative gauge field
  backgrounds.
\newblock {\em Phys. Rev.}, D64:065024, 2001.

\bibitem{Azakov:1997xk}
S.~Azakov.
\newblock The Schwinger model on the torus.
\newblock {\em Fortsch. Phys.}, 45:589--626, 1997.

\bibitem{tata}
D.~Mumford.
\newblock {\em Tata-Lectures about Theta I}.
\newblock Birkh{\"a}user, Boston, 1993.

\end{thebibliography}
\end{document}